\documentclass[a4paper,11pt]{article}
\pdfoutput=1
\usepackage{graphicx}
\usepackage{epsf,amsmath,bbold,amsfonts,stmaryrd}

\usepackage[utf8]{inputenc}
\usepackage{mathrsfs}
\usepackage{appendix}
\usepackage{amssymb}
\usepackage{float}
\usepackage{color}
\usepackage{cite}
\usepackage{hyperref}
\hypersetup{pageanchor=false}
\usepackage{indentfirst}
\usepackage{url}
\usepackage{float}
\usepackage{caption}
\usepackage[numbers,square,comma,sort&compress,merge]{natbib}

\def\mc{\mathcal}

\def\be{\begin{equation}}
\def\ee{\end{equation}}

\def\bea{\begin{eqnarray}}
\def\eea{\end{eqnarray}}

\def\ba{\begin{array}}
\def\ea{\end{array}}

\def\bc{\begin{center}}
\def\ec{\end{center}}

\def\bl{\begin{flushleft}}
\def\el{\end{flushleft}}

\def\br{\begin{flushright}}
\def\er{\end{flushright}}

\def\bi{\begin{itemize}}
\def\ei{\end{itemize}}

\def\bt{\begin{tabular}}
\def\et{\end{tabular}}

\numberwithin{equation}{section}

\begin{document}
\title{\textbf{Innermost stable circular orbit and shadow  of the $4D$ Einstein-Gauss-Bonnet black hole }}
\author{
Minyong Guo$^{1}$ and Peng-Cheng Li$^{1,2*}$}
\date{}
\maketitle

\vspace{-10mm}

\begin{center}
{\it
$^1$Center for High Energy Physics, Peking University,
No.5 Yiheyuan Rd, Beijing 100871, P. R. China\\\vspace{1mm}

$^2$Department of Physics and State Key Laboratory of Nuclear
Physics and Technology, Peking University, No.5 Yiheyuan Rd, Beijing
100871, P.R. China\\\vspace{1mm}
}
\end{center}

\vspace{8mm}

\begin{abstract}
Recently, a novel $4D$ Einstein-Gauss-Bonnet (EGB) gravity was formulated by D. Glavan and C. Lin \cite{Glavan:2019inb}. Although whether the theory is well defined is currently debatable, the spherically symmetric black hole solution is still meaningful and worthy of study.  In this paper, we study the geodesic motions in the spacetime of the spherically symmetric black hole solution.  First of all, we find that a negative GB coupling constant is allowable, as in which case the singular behavior of the black hole can be hidden inside the event horizon. Then we calculate the innermost stable circular orbits (ISCO) for massive particles, which turn out to be monotonic decreasing functions of the GB coupling constant. Furthermore, we study the unstable photon sphere and shadow of the black hole. It is interesting to find that the proposed universal bounds on black hole size in \cite{Lu:2019zxb} recently  can be broken when the GB coupling constant takes a negative value.
\end{abstract}

\vfill{\footnotesize  minyongguo@pku.edu.cn,\,\, lipch2019@pku.edu.cn,\\$~~~~~~*$ Corresponding author.}

\maketitle

\newpage
\section{Introduction}
In classical general relativity, singularity is one of the most fundamental questions. The first version of singularity theorem is proposed by Penrose in 1965 \cite{Penrose:1964wq}, which states that the formation of singularities in spacetime is inevitable assuming the weak energy condition and global hyperbolicity. The singularity theorem that people often refer to is the version presented and proved by Hawking and Penrose in 1970 \cite{Hawking:1969sw}, which says a spacetime $\mathcal{M}$ cannot satisfy causal geodesic completeness provided that Einstein’s equations and some assumptions hold. In contrast to the other singularity theorems, the conditions required by the Hawking-Penrose singularity theorem are the easiest to implement physically and cover a wide range of areas.  Up to now, many attempts have been made to eliminate the singularity, which includes and is not limited to considering quantum corrections \cite{Tomozawa:2011gp,Cognola:2013fva,Zhang:2019dgi} and alternative gravities \cite{Ansoldi:2008jw}. Recently, a novel $4D$ Einstein-Gauss-Bonnet (EGB) gravity was formulated by D. Glavan and C. Lin \cite{Glavan:2019inb}. By focusing on the positive GB coupling constant, they discovered a static and spherically symmetric black hole solution, which is practically free from the singularity problem. It's interesting to note that the same solution was already found before, initially in the gravity with a conformal anomaly \cite{Cai:2009ua} and then in gravity with quantum corrections \cite{Tomozawa:2011gp,Cognola:2013fva}. In contrast, in \cite{Glavan:2019inb}  the GB action should be considered as a classical modified gravity theory, so the theory is on an equal footing with general relativity.

However, since the publication of the paper \cite{Glavan:2019inb}, there have appeared several works \cite{Ai:2020peo, Gurses:2020ofy, Shu:2020cjw, Mahapatra:2020rds, Tian:2020nzb, Bonifacio:2020vbk, Arrechea:2020evj} debating that the procedure of taking $D\to 4$ limit in \cite{Glavan:2019inb} may not be consistent. For example, in \cite{Bonifacio:2020vbk} by studying tree-level graviton scattering amplitudes it was shown that in four dimensions there are no new scattering amplitudes than those of the general gravity. On the other hand, some proposals have been raised to circumvent the issues of the novel $4D$ EGB gravity. These proposals can be divided into two classes. One is adding an extra degree of freedom to the theory. For example,  \cite{Lu:2020iav, Kobayashi:2020wqy} considered using the Kaluza-Klein approach of the $D\to 4$ limit to obtain a well-defined theory that belongs to the Horndeski class \cite{Horndeski:1974wa}. The same theory can also be deduced by introducing a counter term into the action \cite{Hennigar:2020lsl,Fernandes:2020nbq}. The other proposal is to keep the two dynamical  degrees of freedom unchanged  at the price of breaking the
temporal diffeomorphism invariance \cite{Aoki:2020lig}. In summary, the novel $4D$ EGB gravity formulated in \cite{Glavan:2019inb} may run into trouble at the level of action or equations of motion.  Nevertheless, the spherically symmetric black hole solution derived in \cite{Glavan:2019inb} and in early literatures \cite{Cai:2009ua,Tomozawa:2011gp,Cognola:2013fva} can be successfully reproduced in those consistent theories of $4D$ EGB gravity, which is a little bit surprise. Therefore, the spherically symmetric black hole solution itself is meaningful and worthy of study.

It can be expected that due to the publication of \cite{Glavan:2019inb}, lots of works concerning  every aspect of the spherically symmetric $4D$ EGB black hole solution will emerge, including theoretical study and the viability of the solution in the real world. In astronomical survey, the singularities of black holes cannot be directly observed, since they are always inside the event horizons of the black holes. In fact, the event horizon cannot be directly observed by astronomical telescopes. However, the emergence of black hole photograph shows, the black hole shadow and the orbit of the light emitter around the black hole can be seen by the Event Horizon Telescope (EHT), and thus the parameters of a black hole can be identified based on the black hole model \cite{Akiyama:2019cqa,Akiyama:2019eap}. On the other hand, the first detection of gravitational waves from a binary black hole merger by the LIGO/Virgo Collaborations  \cite{Abbott:2016blz} opened a new window to probe gravity in the strong field regime, which then enables us to test gravity theories alternative to general relativity \cite{LIGOScientific:2019fpa}. The progress in both areas may help us to distinguish Schwarzschild black hole from other black hole models, including the $4D$ EGB black hole, in the near future.

Based on these, we would like to investigate the geodesic motions of both timelike and null particles in the background of the $4D$ EGB black hole, by focusing on the innermost stable circular orbit (ISCO) of the timelike particle, the unstable photon sphere and the associated shadow of the black hole. The ISCO plays an important role in the study of realistic astrophysics and gravitational wave physics. For example, in the Novikov-Thorne accretion disk model \cite{Page:1974he}, the inner edge of the disk is at the ISCO. Moreover, according to the Buonanno-Kidder-Lehner approach \cite{Buonanno:2007sv}, one can estimate the final black hole spin of a binary black hole coalescence with arbitrary initial masses and spins. The key point is that one may approximate the merger process as a test timelike particle orbiting at the ISCO around a Kerr black hole.  On the other hand, for the motion of the null particles, besides the observable black hole shadow, the photon sphere (or the light ring) provides information on the quasinormal modes of the final black hole in the ringdown phase of the black hole merger \cite{Cardoso:2008bp} (see however \cite{Konoplya:2017wot}). From the theoretical point of view, a sequence of inequalities were proposed recently in \cite{Lu:2019zxb}, which involve the radii of the event horizon, the photon sphere and the shadow. It would be interesting to verify the conjecture for the $4D$ EGB black hole.

Before we get started, we note in \cite{Glavan:2019inb} the black hole solution is constrained  to the positive GB coupling constant case, i.e.  $\alpha>0$ and leaves a gap for the negative GB coupling constant. Thus, we firstly give a very careful analysis and show the black hole can exist when $\alpha<0$. More precisely, we find that when $-8<\alpha\le1$ there always exists a black hole. In this case the singular behavior of the solution is hidden behind the horizon and outside the horizon the solution is well defined. For the solution, according to the analysis of \cite{Cai:2009ua}, we would like to stress that the black hole entropy has the logarithmic behavior. Then, for the first time, we calculate the radius of the ISCO and give a numerical result for the full range of $\alpha$ \footnote{More recently, the authors in \cite{Konoplya:2020bxa} investigated the stability of the $4D$ EGB black hole via the quasinormal mode. They found that to avoid the eikonal instability \cite{Konoplya:2017ymp} of the gravitational perturbations \cite{Takahashi:2010ye}, the absolute value of the GB coupling has to be relatively small and then they calculated the radius of the shadow in this case. In our work, we will not take this stability issue of the black hole into account for the time being, and let the GB coupling constant be constrained only by the regularity of the metric itself.  }. Also, we obtain an approximate analytical expression when $\alpha$ is very small around $0$. We find the radius of the ISCO in the novel solution can be bigger or smaller than the one in Schwarzschild black hole depending on the value of $\alpha$. For the photon sphere and the shadow, we find the exact expressions not only for $0<\alpha\le 1$ but also for $-8<\alpha<0$. Comparing the result to that of the Schwarzschild black hole, we find the $4D$ EGB black  hole contains more features and information which deserves further study.

The paper is organized as follows. In section \ref{section2}, we revisit the spherically symmetric $4D$ EGB black hole solution and determine the full range of $\alpha$ when the spacetime contains a black hole. In section \ref{section3}, we move to the innermost stable circular orbit of the timelike particle. Next, we turn our attention to the photon sphere and shadow in section \ref{section4}.  Finally, in section \ref{summary}, we summarize the results. In this work, we have set the fundamental constants $c$ and $G$ to unity, and we will work in the convention $(-,+,+,+)$.

\section{Revisit the $4D$ EGB black hole solution}\label{section2}
The Einstein-Hilbert action supplemented by a GB term in $D$ dimensions has the form
\be
I=\frac{1}{16\pi G}\int\sqrt{-g}\, d^Dx\left[R+\alpha\left(R_{\mu\nu\lambda\delta}R^{\mu\nu\lambda
\delta}-4R_{\mu\nu}R^{\mu\nu}+R^2\right)\right],
\ee
where $\alpha$ is the  GB coupling
constant. The static
and spherically symmetric black hole solution in this theory was already found in $D\geq 5$ \cite{Boulware:1985wk}. But in $D=4$, the GB term is a total derivative, and hence does not contribute to the gravitational dynamics, unless an extra scalar filed is introduced to be coupled with the GB term, which is known as Einstein-dilaton-Gauss-Bonnet theory \cite{Kanti:1995vq,Kanti:1997br}. However, recently Glavan and Lin \cite{Glavan:2019inb} found that by rescaling the coupling constant,
\be
\alpha\to\frac{\alpha}{D-4},
\ee
of the GB term, and then consider the limit $D\to4$, the Lovelock' s theorem can be bypassed and there exists a spherically symmetric black hole solution in this case. The solution is found to be
\be\label{solution}
ds^2=-f(r)dt^2+\frac{dr^2}{f(r)}+r^2d\Omega^2,
\ee
with
\be
f(r)=1+\frac{r^2}{2\alpha}\left(1-\sqrt{1+\frac{8\alpha M}{r^3}}\right),
\ee
where $M$ is the mass of the black hole. As we mentioned in the introduction, the above limiting procedure has been called into question by subsequent works \cite{Ai:2020peo, Gurses:2020ofy, Shu:2020cjw, Mahapatra:2020rds, Tian:2020nzb, Bonifacio:2020vbk, Arrechea:2020evj}. Whereas the above solution (\ref{solution}) can be exempted from being meaningless, as which is reproduced from those consistent $4D$ EGB gravity theories \cite{Lu:2020iav, Kobayashi:2020wqy,Hennigar:2020lsl,Fernandes:2020nbq,Aoki:2020lig}. In the following all our discussion will be based on the black hole solution (\ref{solution}), which can be independent of the original gravity theory \cite{Glavan:2019inb}.

For the black hole solution, \cite{Glavan:2019inb} argued that there is no real solution at short radial distance $r^3<-8\alpha M$ if $\alpha<0$, and so only the positive $\alpha$ case was considered. However, we will discuss the range of $\alpha$ in detail and claim that the solution always behaves well outside the (outer) event horizon for $-8<\alpha\le1$. In other words, the singular point $r=-2(\alpha M)^{1/3}$ is always hidden inside the  horizon, and thus the metric function  $f(r)$ is always positive outside the  horizon. Despite that in this case there is no resolution for the singularity problem occurring at $r=0$.

\begin{figure}[h]
\begin{center}
\includegraphics[width=130mm,angle=0]{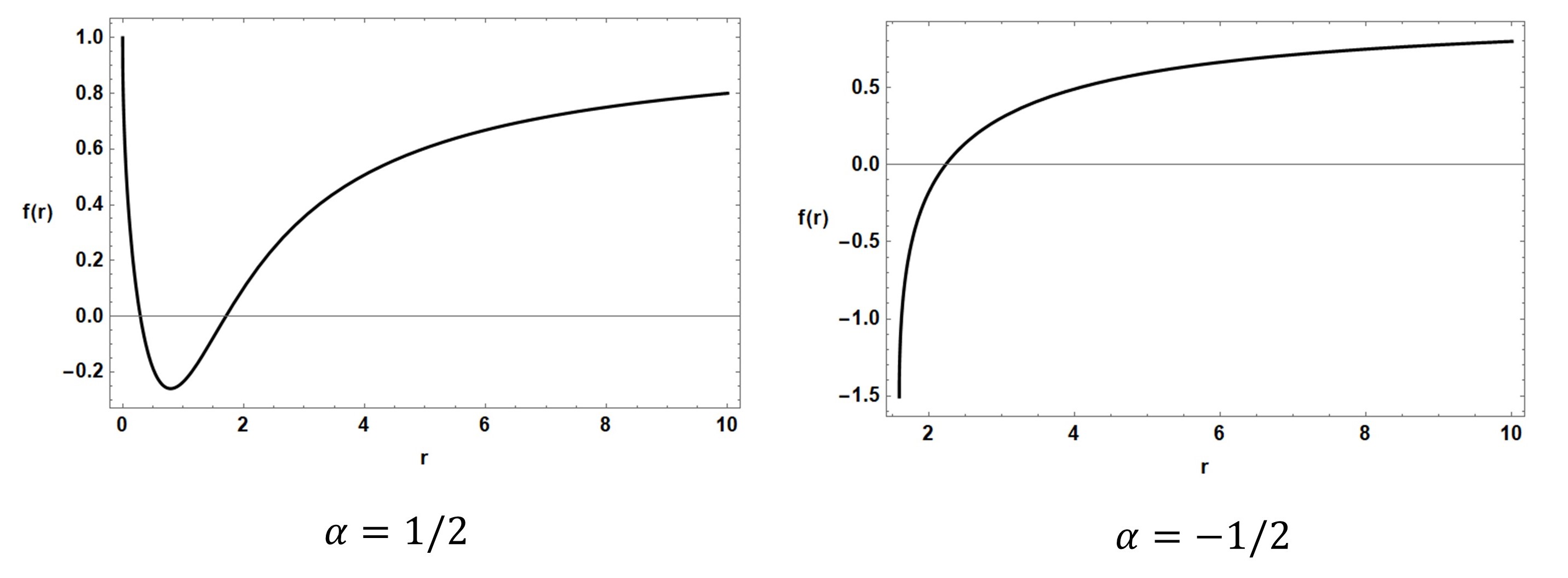}
\end{center}
\vspace{-5mm}
 \caption {The graph of the metric function $f(r)$ with respect to $r$ for two typical values of $\alpha$. }\label{ffunc}
\end{figure}

For simplicity and without loss of generality, we set $M=1$ in the rest of this paper. Let\rq{}s start with the property of the function $f(r)$. From $f^\prime(r)=0$, we find $f(r)$ only has one extreme point at $r=\alpha^{1/3}$. In addition, for $\alpha>0$, we have
\be
f(\infty)=1=f(0^+).
\ee
Thus, $f(\alpha^{1/3})=1-\alpha^{-1/3}$ is the minimum of the function. In order to ensure the existence of horizon we need the condition $f(\alpha^{1/3})\le0$ hold which implies $0<\alpha\le1$, see an example in Fig. \ref{ffunc}, where we take $\alpha=1/2$  for $0<\alpha\le1$ and $\alpha=-1/2$ for $\alpha<0$. In this case, the radii of the horizons read
\be
r_{\pm}=1\pm\sqrt{1-\alpha}.
\ee
While for $\alpha<0$, since $r=\alpha^{1/3}<0$, we have to confine $f(-2(\alpha )^{1/3})=1+2\alpha^{-1/3}<0$ to make sure the existence of the only horizon, which gives us $-8<\alpha<0$. For this situation, we find the single horizon is at
\be
r=1+\sqrt{1-\alpha}.
\ee
Hereto, we have shown the $4D$ EGB black hole exists when $-8<\alpha<0$ and $0\leq\alpha\le1$. And we would like to stress that one shouldn\rq{}t ignore the branch $-8<\alpha<0$ when discussing the whole property of the $4D$ EGB black hole.

Here we use a few words to talk about the  thermodynamic properties of the black hole solution. As we mentioned in the introduction, the solution (\ref{solution}) was initially found in gravity with conformal anomaly \cite{Cai:2009ua}. Therefore, as was studied in that paper, there
exists a logarithmic correction to the well-known Bekenstein-Hawking area entropy. In this case the Wald’s entropy formula \cite{Wald:1993nt} cannot be applied. Instead one can turn to the first law of black hole thermodynamics for help, i.e. $dM=TdS$, with the temperature being
\be
T=\frac{r_h^2-\alpha}{4\pi r_h(r_h^2+2\alpha)},
\ee
where $r_h$ denotes the radius of the event horizon.
The positivity of the temperature requires that the GB coupling constant satisfies the bound which is exactly the same as the one we derived above, i.e. $-8<\alpha\le1$. The entropy is then given by
\be
S=\frac{A}{4}+2\pi\alpha\log{\frac{A}{A_0}},
\ee
where $A=4\pi r_h^2$ is the horizon area and $A_0$ is a constant with dimension of area. For more details on the discussions of the logarithmic behavior of the entropy one can refer to \cite{Cai:2009ua}. More recently, it was demonstrated that if the black hole solution (\ref{solution}) stems from  the consistent theory of the $4D$ EGB gravity which belongs to a class of Horndeski theory, then the black hole entropy can be computed by applying the Wald formula \cite{Lu:2020iav}.
\section{The innermost stable circular orbit of the $4D$ EGB black hole}\label{section3}

In this section, we will calculate the radius of the innermost stable circular orbit for a time particle in the background of the $4D$ EGB black hole. The geodesic motion of a particle is governed by the Hamiltonian
\be
H=\frac{1}{2}g_{\mu\nu}p^\mu p^\nu=-\frac{1}{2}m^2,
\ee
where $m$ is the mass of the particle. $m=0$ describes the null particle and non-zero $m$ corresponds to the timelike particle. Since the $4D$ EGB black hole is a static and spherically symmetric solution, one can always restrict the particle in the equatorial plane, thus the 4-velocity of a timelike particle takes in this form
\bea
\dot{x}^\mu=(\dot{t}, \dot{r}, 0, \dot{\phi}),
\eea
where $\cdot$ represents the derivation of the function with respect to the proper time. Combined with the two conserved quantities for such geodesic, i.e.
\bea
E=-p_t,\quad\quad L=p_\phi,
\eea
then we obtain the orbit equation,
\be\label{orbiteq}
\left(\frac{dr}{d\phi}\right)^2=V_{eff},
\ee
with the effective potential given by
\be
V_{eff}=r^4\left(\frac{E^2}{L^2}-\frac{f(r)}{r^2}-\frac{f(r) m^2}{L^2}\right).
\ee
Circular orbits correspond to $V_{eff}=0$ and $V_{eff}^\prime(r)=0$, where $\prime$ denotes the derivative with respect to the radius $r$. Using these two equations, we have
\bea\label{ee}
e^2&=&\frac{2 f(r)^2}{2f(r)-r f'(r)}=\frac{\left(r^2+2\alpha-r^2\sqrt{\frac{r^3+8\alpha}{r^3}}\right)\left(-r^3-8\alpha+r^3\sqrt{\frac{r^3+8\alpha}{r^3}}+2r\alpha\sqrt{\frac{r^3+8\alpha}{r^3}}\right)}{4\alpha^2\mathcal{R}},\\\label{jj}
j^2&=&\frac{r^3f'(r)}{2f(r)-r f'(r)}=\frac{r^2\left(-r^3-2\alpha+r^3\sqrt{\frac{r^3+8\alpha}{r^3}}\right)}{2\alpha \mathcal{R}},
\eea
where
\be
\mathcal{R}(r)=\left(-3+r\sqrt{\frac{r^3+8\alpha}{r^3}}\right),
\ee
and we have defined
\be
e\equiv \frac{E}{m},\quad\quad j\equiv\frac{L}{m},
\ee
to represent the energy per unit mass and angular momentum per unit mass, respectively.

\begin{figure}[h]
\begin{center}
\includegraphics[width=130mm,angle=0]{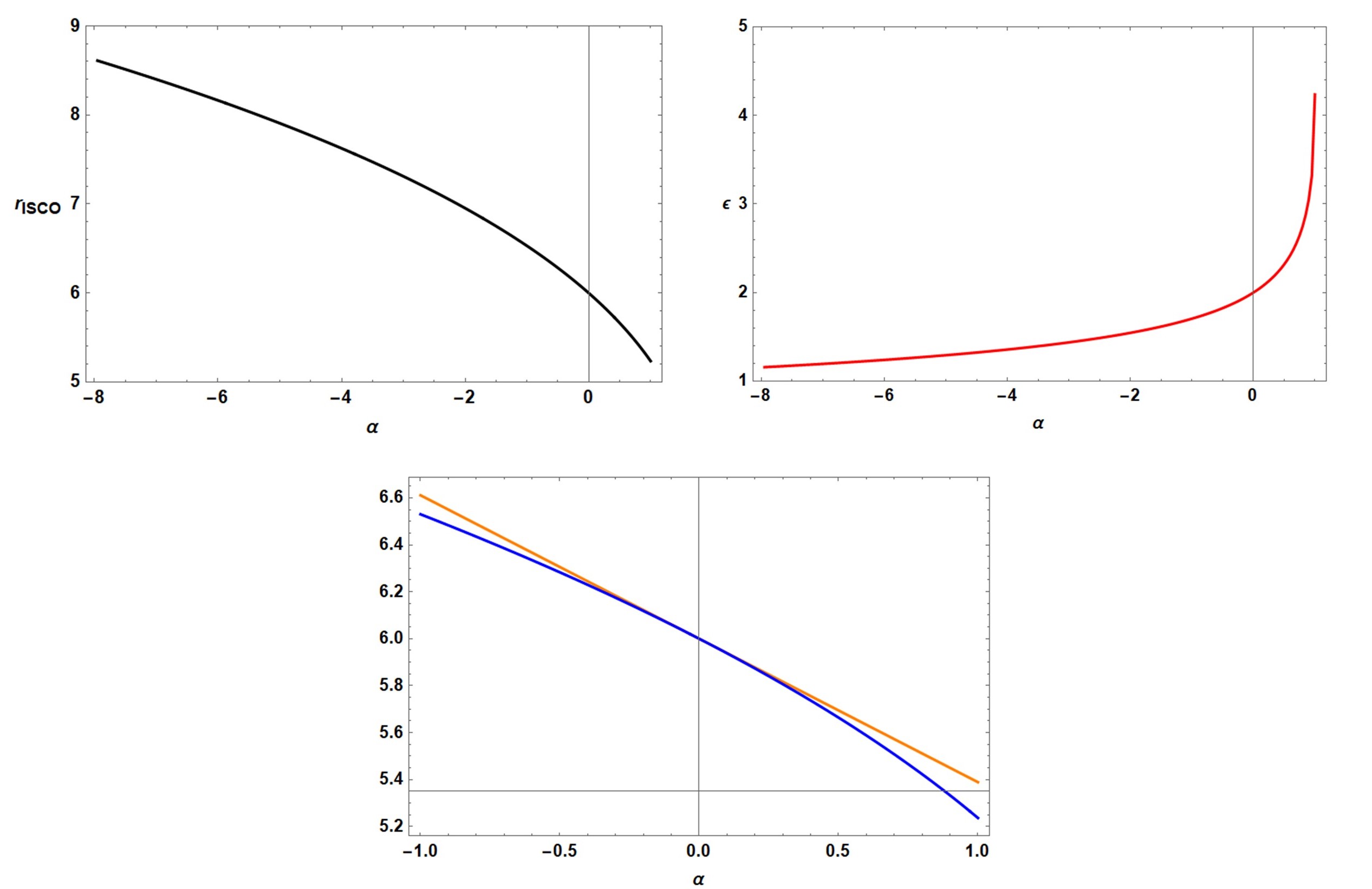}
\end{center}
\vspace{-5mm}
 \caption {The upper left panel shows the dependence of the ISCO radius on the GB coupling constant $\alpha$. The upper right panel shows the deviation of ISCO from the event horizon with respect to $\alpha$. The lower panel is the comparison of numerical result of ISCO radius with the approximate one (\ref{riscoapp}), where the orange line denotes the approximate result and the blue one denotes the numerical result.}\label{isco}
\end{figure}

Circular orbits do not exist for all values of $r$, as the right hand of Eqs. (\ref{ee}) and (\ref{jj}) must be non-negative. Since these expressions are very complex, we prefer to leave it to check in the following discussions. Besides, we observe that the function $\mathcal{R}(r)$ appears in the denominator of Eqs. (\ref{ee}) and (\ref{jj}), then the limiting case of the equality of the two equations gives an orbit with vanishing rest mass, i.e., a photon orbit. As we will see in the next section, the photon sphere is the innermost boundary of the circular orbits for null particles and it occurs at the root of $\mathcal{R}(r)=0$.
In this section, our attention is focused on the ISCO for massive particles, so we leave the discussion of photon sphere and black hole shadow to the next section.

The circular orbits are not all stable. Stability requires that $V_{eff}^{\prime\prime}\le0$ and the equality gives us the location of ISCO. Formally, $r_{ISCO}$ can be calculated from
 \be
 r_{ISCO}=-\frac{3 f( r_{ISCO}) f'( r_{ISCO})}{f( r_{ISCO}) f''( r_{ISCO})-2 f'( r_{ISCO})^2},
 \ee
which works for the general spherically symmetric metric of the form (\ref{solution}). Since the exact expression is not very illuminating, we only  give the numerical result of the radius of the ISCO, see the upper left panel of Fig. \ref{isco}. To keep our result self-consistent, we substitute the radius of the ISCO in the right hand of Eqs. (\ref{ee}) and (\ref{jj}), after some non-trivial algebraic manipulations, it can be shown that when $r\ge r_{ISCO}$, $e^2$ and $j^2$ are always positive.

We find that the radius of the ISCO is a decreasing function of $\alpha$, and so when $0<\alpha\le1$, the ISCO is smaller than the one of the Schwarzschild black hole, i.e. $r_{ISCO}<6$, while $r_{ISCO}>6$ for $-8\le\alpha<0$. The corresponding angular momentum per unit mass $j$ and  energy per unit mass  $e$ decrease with $\alpha$ as well. In addition, we introduce a new parameter
\be
\epsilon=\frac{r_{ISCO}-r_h}{r_h},
\ee
to characterize the extent to which the ISCO radius deviates from the radius of the event horizon. We find that $\epsilon$ is increasing with $\alpha$, as shown in the top right panel of Fig. \ref{isco}. Moreover, when $\alpha$ is very small around $0$, we obtain an approximate analytic result of the ISCO, that is
\be\label{riscoapp}
r_{ISCO}=6-\frac{11}{18}\alpha+\mathcal{O}(\alpha^2),
\ee
with $e=\frac{2 \sqrt{2}}{3}-\frac{\alpha }{162 \sqrt{2}}+\mathcal{O}\left(\alpha ^2\right)$ and
$j=2 \sqrt{3}-\frac{\alpha }{6 \sqrt{3}}+\mathcal{O}\left(\alpha ^2\right)$.
In Fig. \ref{isco}, we also show the comparison between the numerical result with our approximate result. As expected, we find they match very well for a small $\alpha$. This approximate analytical result may be helpful if an astronomical black hole has a very small deviation from the Schwarzschild black hole. Another notable case is $\alpha=1$, which corresponds to the extremal EGB black hole. In this case, we find $e=0.94$, $j=3.35$ and $r_{ISCO}=5.24$.

The effects of the GB coupling constant on the ISCO radius may be reflected in some astronomical phenomena. For example, since a positive $\alpha$ leads to a smaller ISCO and smaller event horizon, which means the merger time of the coalescence of the black hole binary is later than that of the Schwarzschild black holes. As a consequence, the chirp mass and the total mass of the system might be underestimated  when matching with the gravitational wave template based on general relativity. Similar situation occurs for the coalescence of two charged black holes when their charges are of the same kind \cite{Christiansen:2020pnv,Wang:2020fra}.

\section{Photon sphere and shadow of the $4D$ EGB black hole}\label{section4}
In this section, we will discuss the photon sphere and shadow of the $4D$ EGB black hole. In the geometric optics limit, the motion of a photon is treated as a null geodesic. In the background of the $4D$ EGB black hole, the orbit equation for the null geodesics is just Eq.(\ref{orbiteq}) with $m=0$.  By evaluating the equations $V_{eff}=0$ and $V_{eff}'=0$, we obtain the circular null geodesic occurring at
\be\label{rph}
r_{ph}=2 \sqrt{3} \cos \left[\frac{1}{3} \cos ^{-1}\left(-\frac{4 \alpha }{3 \sqrt{3}}\right)\right].
\ee
Due to the spherical symmetry, the photons will fill all the circular orbits to form a so-called photon sphere. One can easily show that the radius of the photon sphere is a decreasing function of the GB coupling constant $\alpha$. The corresponding constant of motion $L^2/E^2$ for this photon sphere is given by
\be\label{bc}
b_c^2=\frac{r_{ph}^2}{f(r_{ph})}=\Bigg(\frac{1}{r_{ph}^2}-\frac{\sqrt{\frac{8 \alpha }{r_{ph}^3}+1}-1}{2 \alpha }\Bigg)^{-1},
\ee
where $b_c$ is sometimes called the critical impact parameter. One can show that this is a monotonic increasing function of $\alpha$ and when $\alpha$ is small one obtains $b_c=3 \sqrt{3}-\frac{2 \alpha }{3 \sqrt{3}}+\mathcal{O}\left(\alpha ^2\right)$. Moreover, one can check that for arbitrary values taken in the interval $1\geq\alpha>-8$, one always has $V_{eff}''>0$ with $r=r_{ph}$ and $L/E=b_c$, which means the photon sphere in the background of the $4D$ EGB black hole is unstable.
\begin{figure}[h]
\begin{center}
\includegraphics[width=80mm,angle=0]{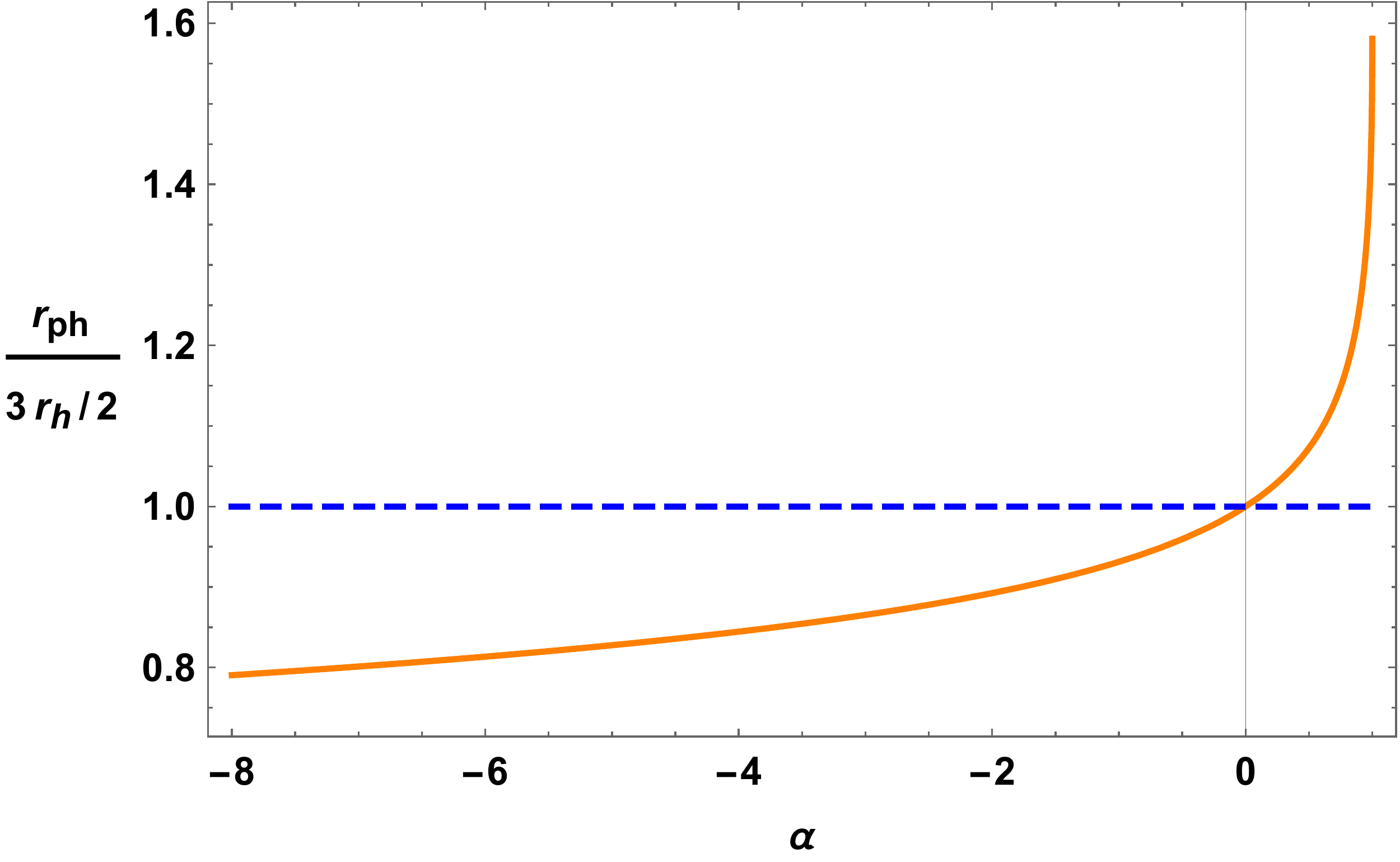}\,
\includegraphics[width=80mm,angle=0]{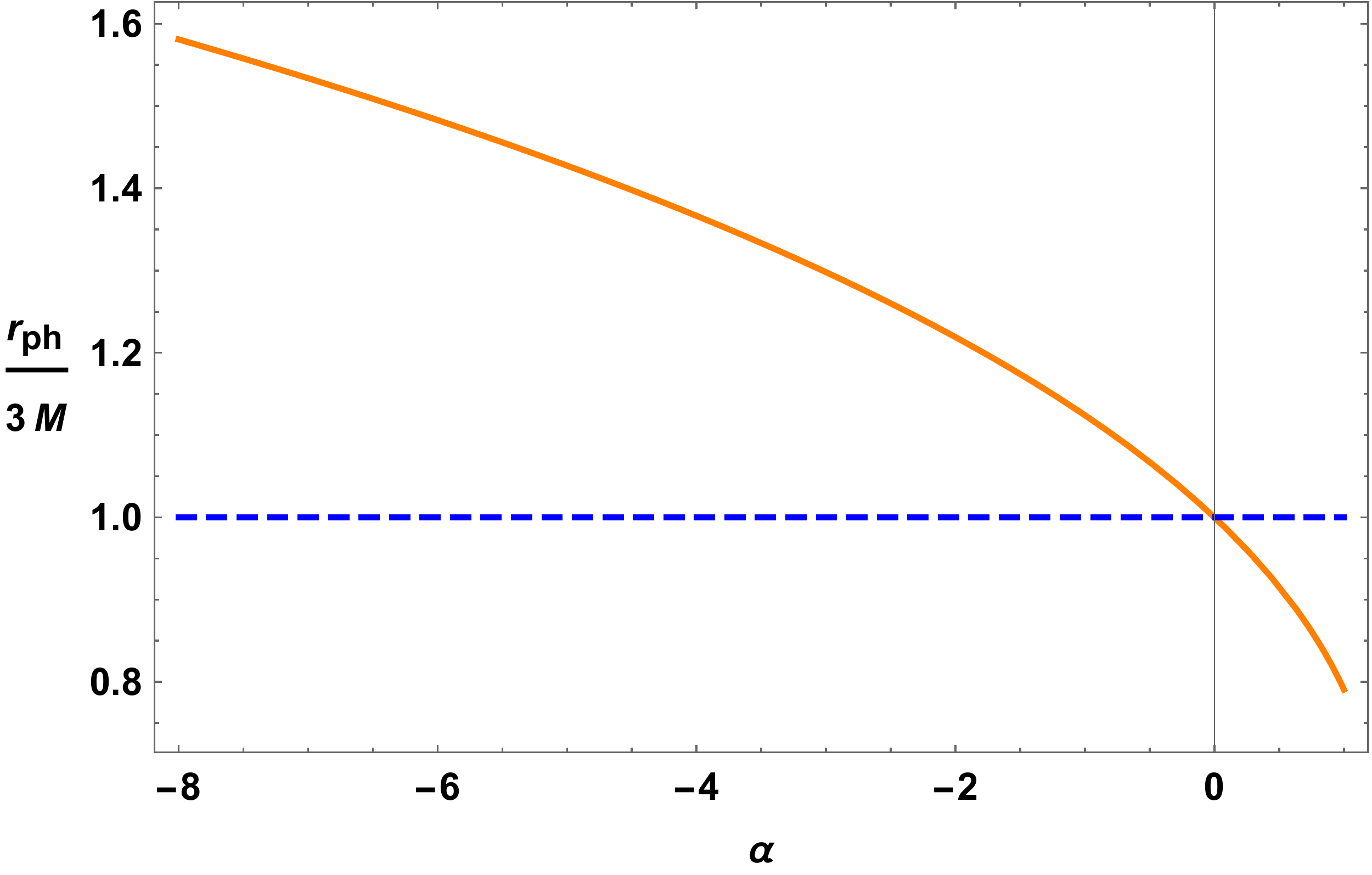}
\end{center}
\vspace{-5mm}
 \caption {the dependence of  $\frac{r_{ph}}{3r_h/2}$ and $\frac{r_{ph}}{3M}$ on the GB coupling constant $\alpha$.}\label{rph-rh3M}
\end{figure}

One interesting property of the $4D$ EGB black hole is that the bounds on the photon sphere proposed by \cite{Lu:2019zxb} can be broken when $\alpha$ is allowed to be negative. In \cite{Lu:2019zxb} \footnote{During the subsequent developments, the study was generalized to the rotating black holes \cite{Feng:2019zzn} and charged EGB black hole in $D\geq5$ dimensions \cite{Ma:2019ybz}, and the conjecture for static black holes in Einstein gravity was proven in \cite{Yang:2019zcn}. }, the authors made a conjecture for a sequence of inequalities involving several parameters characterizing the black hole size, viz.,
\be\label{bounds}
\frac{3}{2}r_h\leq r_{ph}\leq \frac{r_{sh}}{\sqrt{3}}\leq 3M,
\ee
where $r_{sh}$ denotes the radius of the shadow. In what follows we will show that these relations can be violated for the $4D$ EGB black hole. We first focus on the photon sphere and later on turn to the shadow. From Fig. \ref{rph-rh3M}, we can see that when $\alpha\geq0$ the above inequalities works, but when $\alpha<0$, $r_{ph}$ can be less than $3r_h/2$ and $r_{ph}$ can be larger than $3M$. Therefore, in the case
$\alpha\leq0$ the inequalities involving $r_{ph}$ modifies as
\be\label{rhrph}
\frac{3}{2}r_h\geq r_{ph}\geq 3M.
\ee
The existence of unstable photon sphere means the appearance of the observable of the black hole, the black hole shadow. We consider all null geodesics that go from the position of the static observer at $(t_O, r_O, \theta=\pi/2, \phi_O=0)$ into the past. Those critical null geodesics that orbit around the black hole on the photon sphere will leave the observer at an angle $\theta$ with respect to the radial line that satisfies
\be
\tan \theta=\frac{r d\phi}{g_{rr} dr}\Big|_{r=r_O}.
\ee
This angle describes the angular size of the shadow of the black hole.
From the orbit equation (\ref{orbiteq}) we then find
\be
\tan^2 \theta=\frac{f(r_O)}{r_O^2\left(\frac{f(r_{ph})}{r_{ph}^2}-\frac{f(r_{O})}{r_{O}^2}\right)}.
\ee
For a static observer at large distance, i.e., $r_O\gg r_h$, this expression can be further simplified as
\be
\tan\theta\simeq\frac{r_{ph}}{r_O\sqrt{ f(r_{ph})}}.
\ee
Therefore the linear radius of the shadow is simply given by
\be
r_{sh}=r_O \sin \theta\simeq\frac{r_{ph}}{\sqrt{ f(r_{ph})}},
\ee
where $r_{ph}$ is obtained in (\ref{rph}). Clearly, the radius of the shadow is equal to the critical impact parameter $b_c$ (\ref{bc}). In fact, one can find that this is a universal result as long as the metric has the form (\ref{solution}) and the spacetime is asymptotically flat. Since the explicit expression is not very illuminating, so we will not present it here. The same as $b_c$, $r_{sh}$  is a decreasing function of $\alpha$ and up to linear order in $\alpha$, $r_{sh}=3\sqrt{3}-2\alpha/3\sqrt{3}+\mc O(\alpha^2)$. \footnote{Note that this looks different from the result in \cite{Konoplya:2020bxa}, as in their convention the event horizon is set to unity.} Therefore, if the shadow size is measured larger or smaller than the prediction from Schwarzschild black hole, this may be attributed to the result of EGB black hole with a negative GB coupling or positive GB coupling. However, the observational result in the real world in general depends on
many  parameters describing the environment. So one cannot simply connect the result with the EGB black hole.
\begin{figure}[h]
\begin{center}
\includegraphics[width=80mm,angle=0]{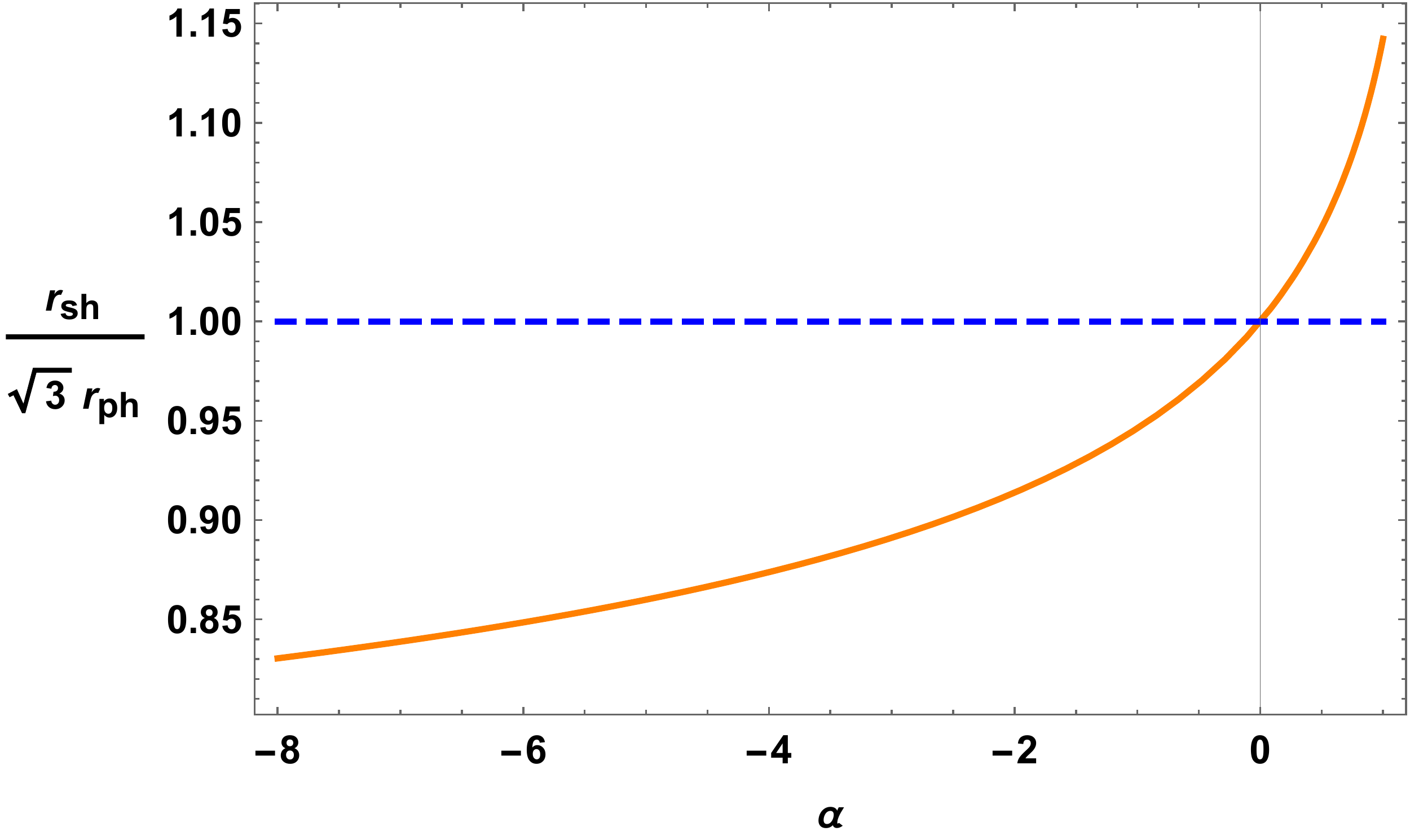}\,
\includegraphics[width=80mm,angle=0]{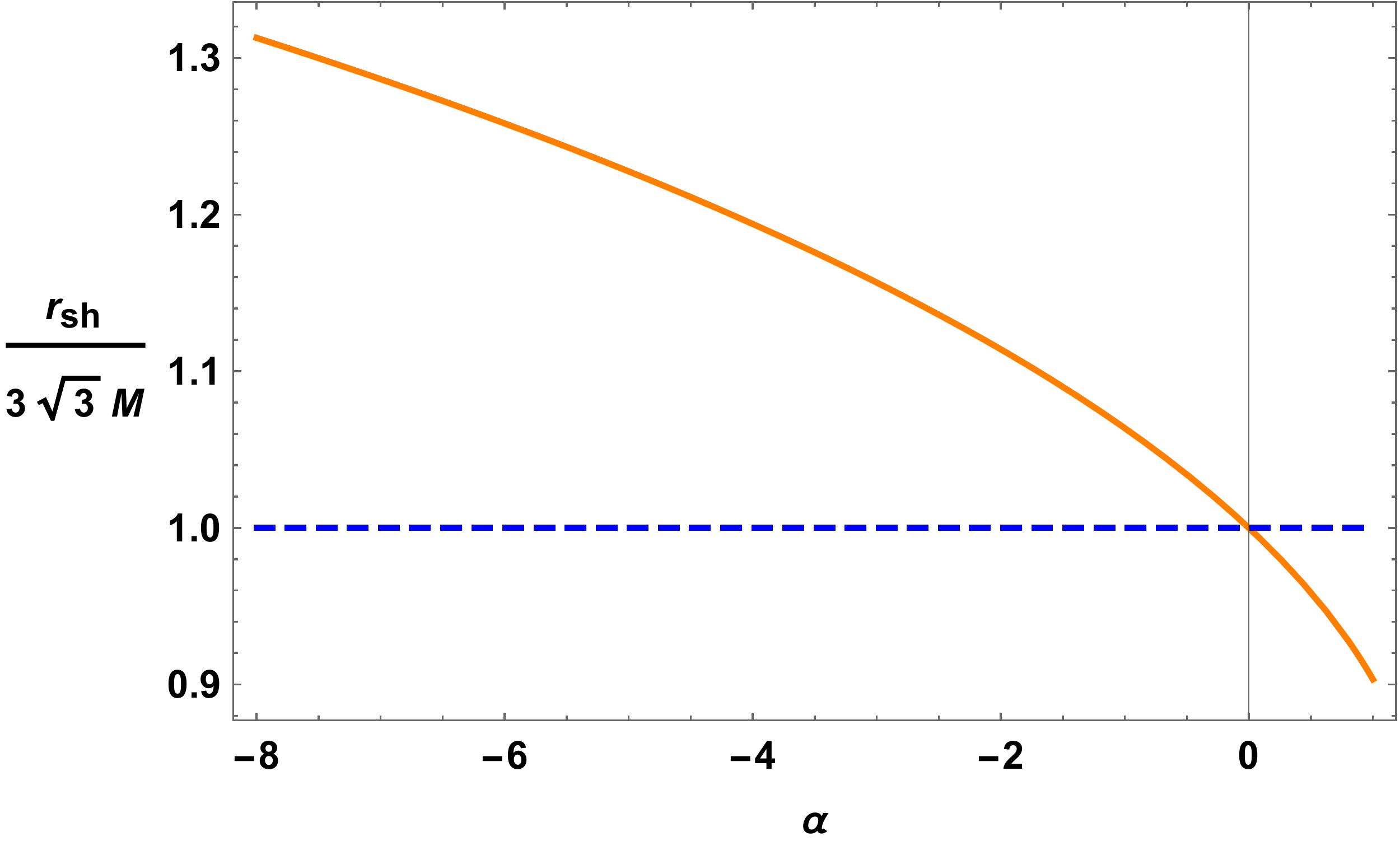}
\end{center}
\vspace{-5mm}
 \caption {the dependence of $\frac{r_{sh}}{\sqrt{3}r_{ph}}$ and $\frac{r_{sh}}{\sqrt{3}M}$ on the GB coupling constant $\alpha$.} \label{rsh-rph3M}
\end{figure}

So far, we find that all the four parameters that characterizing the size of a black hole, including the event horizon, ISCO, photon and shadow, are decreasing functions of the GB coupling constant $\alpha$. Let's now return to the inequalities involving the shadow radius. As is shown in Fig. \ref{rsh-rph3M}, the relations involving $r_{sh}$ obey the inequalities (\ref{bounds}) for a positive $\alpha$, however, for a negative $\alpha$, the relations are reversed, that is,
\be
r_{ph}\geq \frac{r_{sh}}{\sqrt{3}}\geq 3M.
\ee
Combined above inequalities with (\ref{rhrph}), we find that for $\alpha\leq0$, the inequalities (\ref{bounds}) should be totally reversed.
Actually, one can check that for higher dimensional EGB black holes \cite{Ma:2019ybz}, the negative GB coupling constant will lead to the broken of the higher dimensional version of the bounds (\ref{bounds}) as well. The physics behind the broken of the inequalities (\ref{bounds}) is as follows. According to the analysis in \cite{Yang:2019zcn} and \cite{Ma:2019ybz}, a necessary condition for the validity of the inequalities (\ref{bounds}) is that the weak energy condition has to be obeyed. From the view of Einstein gravity, if the GB term is regarded as matter field, the weak energy condition holds for a positive GB coupling but is violated for a negative one.
\section{Summary}\label{summary}
In this paper, we studied the geodesic motions of timelike and null particles in the spacetime of the spherically symmetric $4D$ EGB black hole. We carefully analyzed the metric and found that the GB coupling constant could be negative, because even in this case the singular behavior of the black hole only occurs behind the event horizon. With this extension, we calculated the radius of the innermost stable circular orbit (ISCO) for the timelike particle and found that this radius is a deceasing function of the GB coupling constant. In addition, we calculated the radius of the photon sphere and the angular radius of the shadow of the $4D$ EGB black hole. Besides the ISCO radius, all the other three parameters characterizing the size of the black hole, namely the event horizon, the photon sphere and  the shadow, are decreasing functions of the GB coupling constant, when the mass of the black hole is fixed as unity. As a consequence, the universal bounds on the size of a spherically symmetric black hole proposed in \cite{Lu:2019zxb} can be broken for a negative GB coupling constant.
\section*{Acknowledgments}
We would like to thank Rong-Gen Cai for drawing our attention to the Ref.\cite{Cai:2009ua} and the useful discussion with Aimeric Colléaux.
The work is in part supported by NSFC Grant No. 11335012, No. 11325522, No. 11735001 and No. 11847241. MG and PCL are also supported by NSFC Grant No. 11947210. And MG is also funded by China National Postdoctoral Innovation Program 2019M660278.

\end{document}